\documentclass{blois}
\pdfoutput=1
\usepackage{amsmath}
\usepackage{amssymb}
\usepackage{subfig}

\newcommand{\Photo}{}

\title{DOUBLE HIGGS BOSON PRODUCTION IN THE STANDARD MODEL WITH EXTRA SCALAR
PARTICLES}

\author{E. V. ZHEMCHUGOV}
\address{
 Institute for Theoretical and Experimental Physics, 117218, Moscow, Russia \\
 Moscow Engineering Physics Institute, 115409, Moscow, Russia
}

\newcommand{\Br}{\mathcal{B}}

\def\figscale{0.29}

\begin{document}

\maketitle
\abstracts{
Three extensions of the scalar sector of the Standard Model are considered: one
extra isosinglet, one extra isotriplet, two extra isotriplets (the
Georgi-Machacek model). Double Higgs boson production cross section is
calculated in all these extensions. Bounds from electroweak precision
observables, signal strength measurements and custodial symmetry violation are
estimated. \\[1em]
This paper is a contribution to the proceedings of the 27th Rencontres de
Blois.
}

\section{Introduction}

In 2012 a scalar particle with the mass of 125~GeV has been found at the
LHC.~\cite{atlas-higgs,cms-higgs} In order to confirm that this is the Standard
Model Higgs boson, its couplings have to be measured. Among these couplings, one
of the most interesting is the triple boson coupling which is, up to a
conventional coefficient, equal to the following ratio:
\begin{equation}
 g_{hhh} \sim \frac{m_h^2}{v},
\end{equation}
where $m_h$ is the Higgs boson mass, and $v$ is the vacuum expectation value. We
know the vacuum expectation value with good precision from the Fermi coupling in
muon decay, and now we have measured the higgs mass. Thus, any difference in the
triple coupling constant from the theoretically predicted value would speak of
New Physics in the scalar sector.

The triple coupling constant can be measured in the double higgs production
process. However, the Standard Model prediction for such a process is very small,
just 40~fb at the center-of-mass energy 14~TeV.~\cite{hh-nnlo} Such value can
only be measured at the HL-LHC. But if there indeed is New Physics, it might
increase the double higgs production cross section, and we may be able to
observe it during the Run~2. In this paper three extensions of the scalar sector
of the Standard Model that might provide such an increase are considered.

\section{Isosinglet}

First, let us consider a model with an extra isosinglet.~\cite{hh-isosinglet}
The extended scalar sector consists of two fields:
\begin{equation}
  \Phi = \begin{pmatrix}
   \phi^+ \\ \frac{1}{\sqrt{2}} (v_\Phi + \phi + i \eta)
  \end{pmatrix},
  \  X = v_X + \chi,
\end{equation}
where $\Phi$ is the same isodoublet as in the SM, and $X$ is the new isosinglet.
Both fields have their own vacuum expectation value, $v_\Phi$ and $v_X$, and
$\phi$ and $\chi$ are the two neutral scalar particles. Two additional terms
appear in the potential:
\begin{equation}
 V_1(\Phi, X)
 = -\frac{1}{2} m_\Phi^2 \Phi^\dagger \Phi
 +  \frac{\lambda}{2} (\Phi^\dagger \Phi)^2
 +  \frac{1}{2} m_X X^2
 +  \mu \Phi^\dagger \Phi X.
\end{equation}
They describe the bare mass of the isosinglet, and the mixing of the neutral
particles. There are more terms allowed by Lorentz invariance, but we assume
that they are multiplied by small coupling constants. We introduce the mixing
angle $\alpha$,
\begin{equation}
   \begin{pmatrix}
    h \\ H
   \end{pmatrix}
 = \begin{pmatrix}
    \cos \alpha  & \sin \alpha \\
    -\sin \alpha & \cos \alpha \\ 
   \end{pmatrix}
   \begin{pmatrix}
    \phi \\ \chi
   \end{pmatrix},
 \label{mixing}
\end{equation}
where $h$ and $H$ are the physical eigenstates of the neutral scalar particles.

In total, there are six parameters in the lagrangian: two vacuum expectation
values $v_\Phi$ and $v_X$, two bare masses $m_\Phi$ and $m_X$, and constants
$\lambda$ and $\mu$. Four of them are fixed: (1, 2) for two fields, we get two
equations describing the minimum of the potential; (3) since the isosinglet does
not couple to fermions, we get $v_\Phi$ from the muon decay just as in the SM;
(4) we assume that it is $h$ that was discovered at the LHC, so we set $m_h =
125$~GeV. We are left with two free parameters, and we choose them to be $\sin
\alpha$ and the mass of the second boson,~$m_H$.

$H$ decay widths are just like those of the SM higgs, except that they are
multiplied by $\sin^2 \alpha$. In addition, a brand new decay mode appears:
\begin{equation}
 \Gamma(H \to hh)
  = \frac{(2 m_h^2 + m_H^2)^2}{32 \pi v_\Phi^2 m_H}
    \sin^2 \alpha \cos^4 \alpha
    \sqrt{1 - \left( \frac{2 m_h}{m_H} \right)^2}.
\end{equation}

$H$ production cross section is the same as the SM higgs production cross
section, times $\sin^2 \alpha$. In order to compute the double $h$ production,
we multiply it by the corresponding branching ratio:
\begin{equation}
 \sigma(pp \to H \to hh)
 = \sigma(pp \to h)_\text{SM} \cdot \sin^2 \alpha \cdot \Br(H \to hh).
\end{equation}

Experimentalists provide us with measurements of the following values:
\begin{equation}
 \mu_i = \frac{\sigma(pp \to h) \cdot \Br(h \to f_i)}
              {(\sigma(pp \to h) \cdot \Br(h \to f_i))_\text{SM}},
\end{equation}
where $h \to f_i$ describe different decay modes. In the isosinglet model $\mu_i
= \cos^2 \alpha$. Experimental values combined into a single quantity are:
\begin{equation}
 \begin{aligned}
  \mu &= 1.30_{-0.17}^{+0.18}
   \text{ by the ATLAS collaboration;~\cite{mu-atlas}}
  \\
  \mu &= 1.00_{-0.13}^{+0.14}
   \text{ by the CMS collaboration.~\cite{mu-cms}}
 \end{aligned}
 \label{mu}
\end{equation}
Experiment data prefer values over unity, however values below one are still
allowed at two-sigma level for ATLAS and one sigma for CMS.

\begin{figure}
 \centering
 \begin{minipage}[t]{0.5\textwidth}
  \raisebox{-\height}{\includegraphics[scale=\figscale]{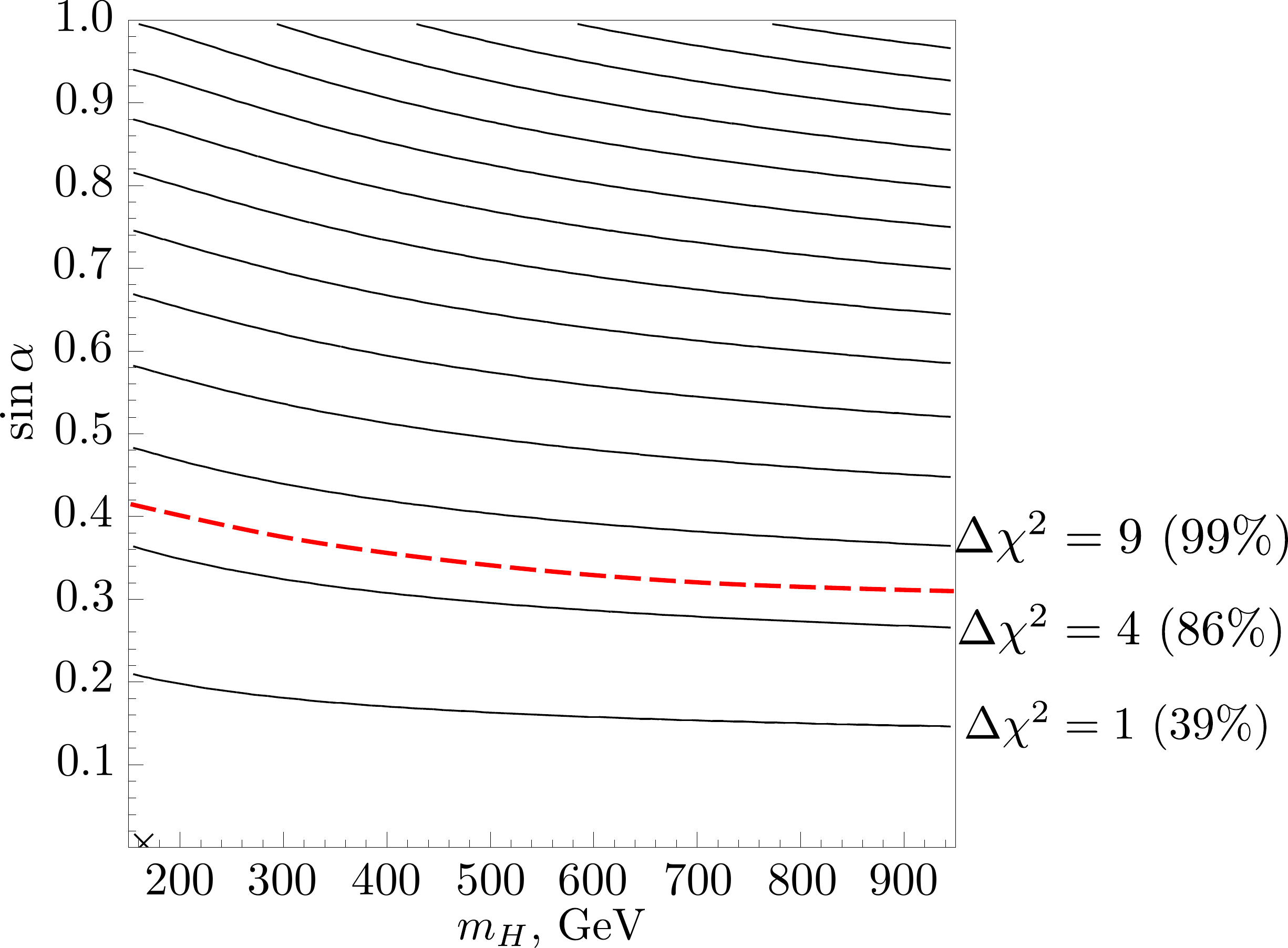}}
 \end{minipage}
 \begin{minipage}[t]{0.45\textwidth}
  \caption{
   Fit of the isosinglet model to electroweak observables and the measurements
   of $\mu$~\eqref{mu}. Values in parentheses are probabilities that numerical
   values of $(m_H, \sin \alpha)$ are below the corresponding line. The dashed
   line corresponds to $\Delta \chi^2 = 5.99$ with the probability 95\%.
  }
  \label{fig:fit}
 \end{minipage}
\end{figure}

To take experimental values into account in a more robust way we have
calculated a fit of electroweak observables and the measurements of $\mu$. The
fit was calculated with the help of the LEPTOP program.~\cite{leptop} Fit results are
presented in fig.~\ref{fig:fit}. The minimum of $\chi^2$ is reached at line
$\sin \alpha = 0$, with $\chi^2 = 19.6$ for the 13 degrees of freedom listed in
our paper.~\cite{hh-isosinglet} From the fit it follows that $\sin \alpha$
cannot be large, with the maximum value of about $0.35$ for confidence
probability 95\% and $m_H = 300$~GeV.

\begin{figure}
 \centering
 \begin{minipage}[t]{0.45\textwidth}
  \includegraphics[scale=\figscale]{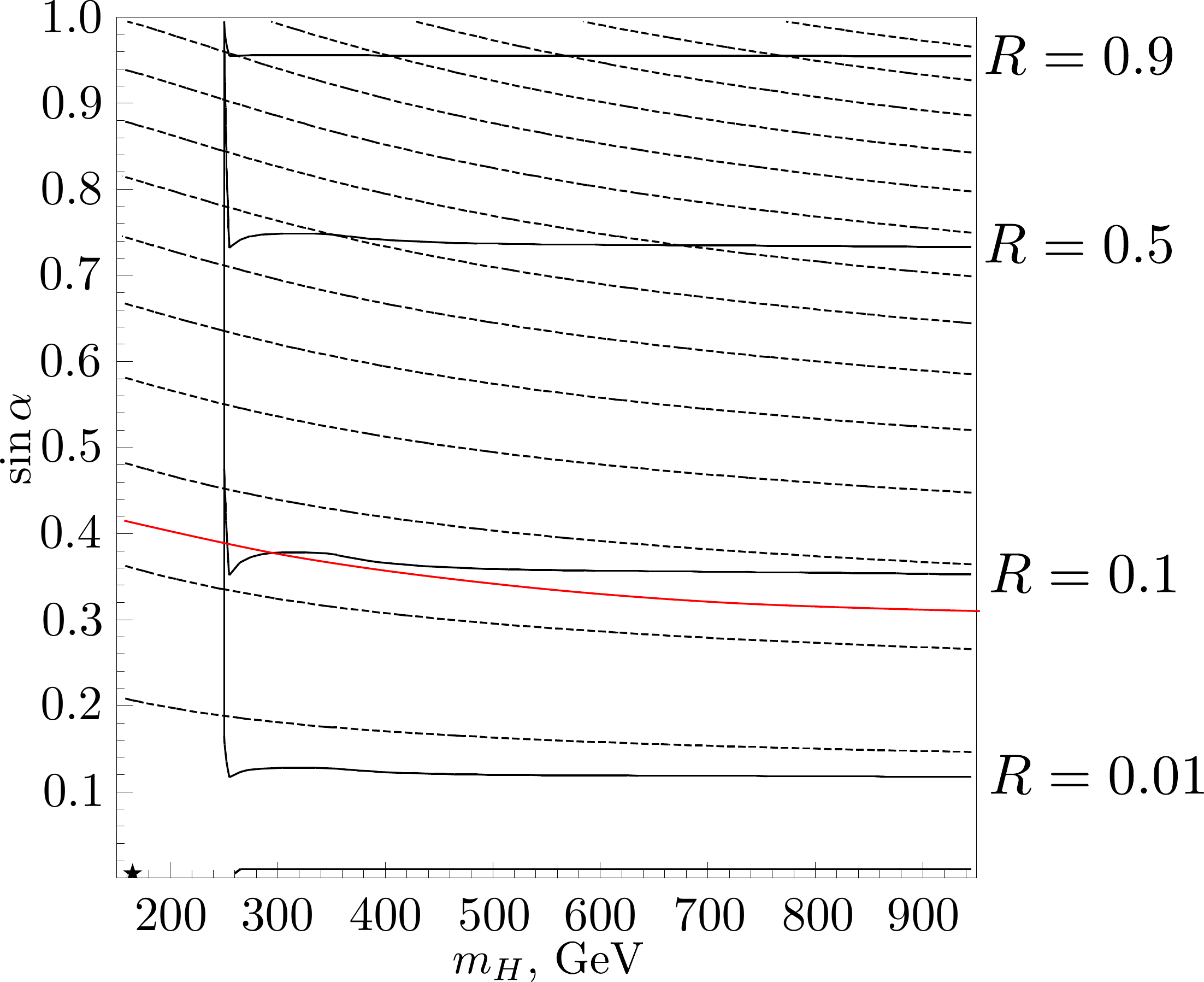}
  \caption{Contour plot of R~\eqref{R}.}
  \label{fig:R}
 \end{minipage}
 \begin{minipage}[t]{0.45\textwidth}
  \includegraphics[scale=\figscale]{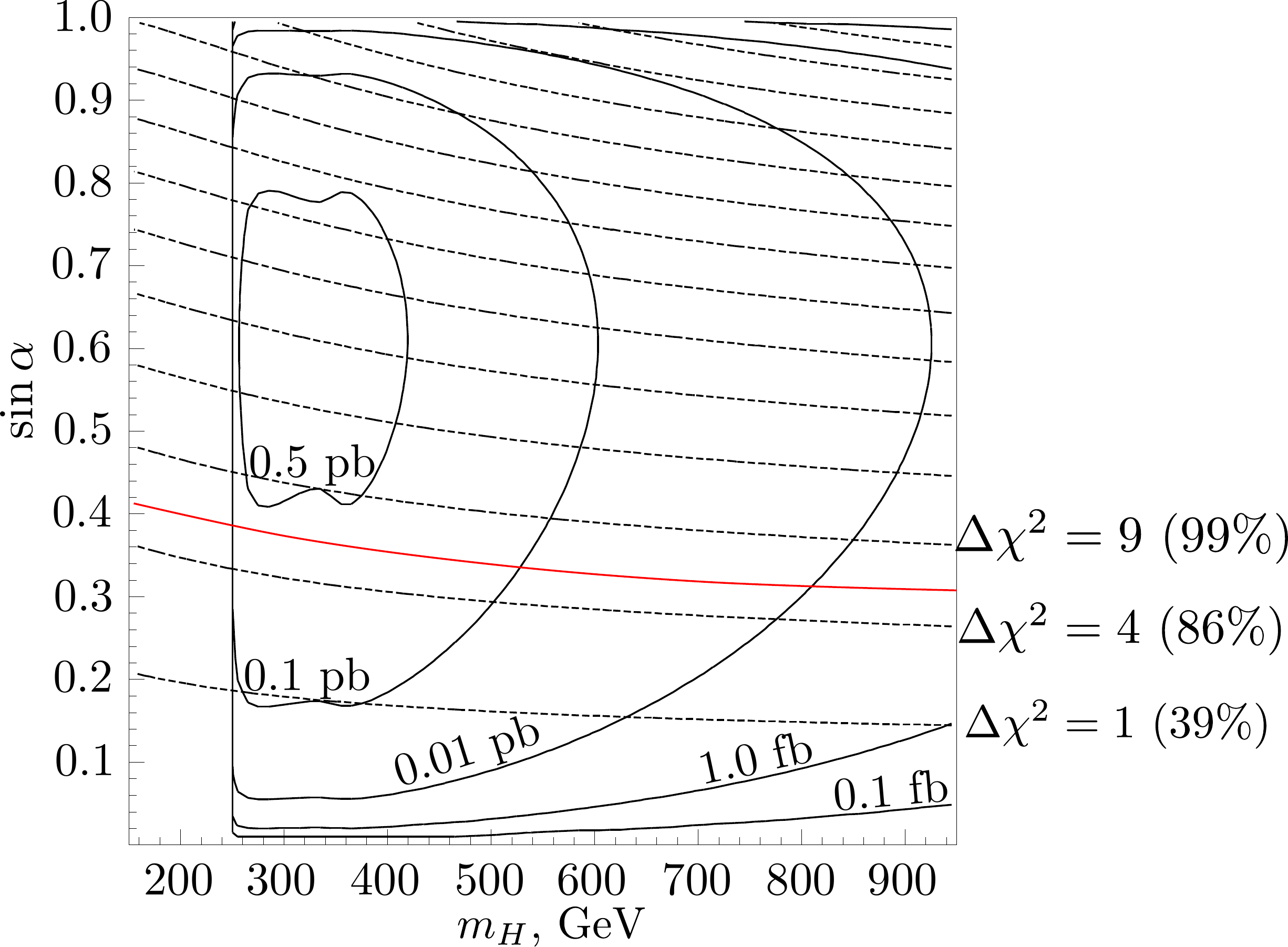}
  \caption{Contour plot of $\sigma(pp \to H \to hh)$ for $\sqrt{s} = 14$~TeV.}
  \label{fig:xsection}
 \end{minipage}
\end{figure}

The golden mode for the search of the new heavy higgs boson is the $H \to ZZ$
decay mode, just as it was for the Standard Model higgs. The expected signal
strength 
\begin{equation}
 R
 \equiv \frac{\sigma(pp \to H) \Br(H \to ZZ)}
             {(\sigma(pp \to h) \Br(h \to ZZ))_\text{SM}}
 = \frac{\sin^4 \alpha}
        {\sin^2 \alpha + \frac{\Gamma(H \to hh)}{\Gamma_\text{SM}}}.
 \label{R}
\end{equation}
Note that it does not depend on collision energy. Contour lines of $R(\sin
\alpha, m_H)$ are presented in fig.~\ref{fig:R}.  For the allowed region of $\sin
\alpha < 0.35$ we get $R < 0.1$. Experimental data has not set bounds on that
level, with only some tension being observed in e.g., CMS
paper,~\cite{cms-zz-signal} fig.~5 at $m_H \approx 250$~GeV and $m_H \approx
300$~GeV.

Double higgs production cross section for the center-of-mass energy 14~TeV is
shown in fig.~\ref{fig:xsection}. In the allowed region it varies from about
0.4~pb at $m_H = 300$~GeV down to the order of several fb as $m_H$ reaches 1~TeV.

\section{Isotriplet}

In the isotriplet model~\cite{hh-isotriplet} the extra fields are conventionally represented
by a $2 \times 2$ matrix:
\begin{equation}
 \Delta
 = \frac{\vec \Delta \vec \sigma}{\sqrt{2}}
 = \begin{bmatrix}
  \delta^+ / \sqrt{2} & \delta^{++} \\
  \frac{1}{\sqrt{2}} (v_\Delta + \delta + i \eta) & -\delta^+ / \sqrt{2}
 \end{bmatrix},
\end{equation}
where $\delta$ is the new neutral particle. The isotriplet also has its own
vacuum expectation value~$v_\Delta$. The other fields ($\eta$, $\delta^+$,
$\delta^{++}$) are of no interest to us at this moment.
We will use the same notation for physical eigenstates as in the case of the
isosinglet~\eqref{mixing} (with $\chi$ replaced with $\delta$).

In contrast to the isosinglet, the isotriplet couples to gauge bosons.
Consequently, its vacuum expectation value produces contributions to masses of
gauge bosons:
\begin{equation}
 m_W^2 = \tfrac{g^2}{4} (v_\Phi^2 + 2 v_\Delta^2), \ 
 m_Z^2 = \tfrac{\bar g^2}{4} (v_\Phi^2 + 4 v_\Delta^2),
\end{equation}
where $g$ is the $SU(2)_L$ coupling and $\bar g = g / \cos \theta_W$, $\theta_W$
is the Weinberg angle. The fact that gauge boson masses are changed nonuniformly
breaks custodial symmetry of the model. The breaking is characterized by the
quantity
\begin{equation}
 \rho
 \equiv  \frac{m_W}{m_Z \cos \theta_W}
 \approx \left( \frac{m_W}{m_Z \cos \theta_W} \right)_\text{SM}
         \left( 1 - \frac{v_\Delta^2}{v_\Phi^2} \right).
 \label{rho}
\end{equation}
The value of $\rho$ provided by
PDG~\cite{pdg}
is $1.00040 \pm 0.00024$. Although it is
greater than 1, we can set the bound $v_\Delta \lesssim 5$~GeV at $3 \sigma$
level, and this is the strongest bound in the isotriplet model. To estimate the
upper bound on the cross section, we will use the value $v_\Delta = 5$~GeV in
following.

From the Fermi coupling constant we get that $v_\Phi \approx 246$~GeV, just like
in the SM. It follows that $\sin \alpha \approx 2 v_\Delta / v_\Phi \approx
1/25$, so only one model parameter remains which is $m_H$. We will consider the
case of $m_H = 300$~GeV so that $H$ has enough mass to decay to $hh$, but not to
$t \bar t$.

It is a peculiar property of this model that $H \to WW$ decay is suppressed as
$(m_h / m_H)^4$, so $H \to ZZ$ is the ``golden mode'' for the heavy Higgs boson
discovery. As for the $H \to hh$ decay, its branching ratio approximately equals
0.8 for the chosen case of $m_H = 300$~GeV.

The SM Higgs boson is produced at the LHC through the six main channels: $gg$,
$WW$ and $ZZ$ fusions, $t \bar t$, $W$ and $Z$ associated productions. Same
is true for the heavy Higgs boson of the isotriplet model. Corresponding cross
sections can be calculated from cross sections of the SM with $(m_h)_\text{SM} =
m_H$. Noting that $gg$ fusion and $t \bar t$ associated production share the
same higgs vertex, we get
\begin{align}
    \frac{\sigma(gg \to H)}{\sigma(gg \to h)_\text{SM}}
  = \frac{\sigma(gg \to H t \bar t)}{\sigma(gg \to h t \bar t)_\text{SM}}
 &= 2.4 \cdot 10^{-3},
 \\
\intertext{Similarly,}
    \frac{\sigma(ZZ \to H)}{\sigma(ZZ \to h)_\text{SM}}
  = \frac{\sigma(Z^* \to ZH)}{\sigma(Z^* \to Zh)}_\text{SM}
 &= 1.0 \cdot 10^{-3},
 \\
    \frac{\sigma(WW \to H)}{\sigma(WW \to h)_\text{SM}}
  = \frac{\sigma(W^* \to WH)}{\sigma(W^* \to Wh)}_\text{SM}
 &= 7.3 \cdot 10^{-5}.
\end{align}
In the SM the gluon fusion channel dominates, being an order of magnitude
greater than the second biggest one, $WW$ fusion. With these numbers it is clear
that gluon fusion dominates even stronger for the heavy Higgs boson of the
isotriplet model. For $\sqrt{s} = 14$~TeV we get $\sigma(gg \to H) = 25$~fb, so
the 125~GeV Higgs boson production cross section gets enhanced to
\begin{equation}
 \sigma(pp \to hh) = 40\text{ fb (SM)} + 25\text{ fb ($H$ production)} \cdot 0.8
 \text{ (branching)} = 60\text{ fb}.
\end{equation}

Custodial symmetry can be saved through introduction of yet another, real
isotriplet, with its vacuum expectation value equal to $v_\Delta$. The
corresponding model is referred to as the Georgi-Machacek
model.~\cite{georgi-machacek} In this case the bound coming from~\eqref{rho} is
removed, and signal strength measurements allow $v_\Delta$ to reach 50~GeV.
Double higgs production cross section can then be as high as 2~pb.

An interesting property of the Georgi-Machacek model is that when the mixing
angle is small, decays of $H$ to vector bosons are severely
suppresed,~\cite{gm-h-vv} with about 98\% decays going through the $H \to hh$
channel for $m_H$ near 300~GeV. Hence, search in the $ZZ$ final mode at the LHC
will not lead to new limits on model parameters.

\section{Conclusions}

Significant enhancement of the cross section for double production of 125~GeV
Higgs bosons can be observed in the isosinglet model. Depending on model
parameters, it can be as high as 0.4~pb for the collision energy of 14~TeV, an
order of magnitude greater than the SM value of 40~fb. Primary model constraints
are set by the signal strength measurements, with experiment data right now
becoming sensitive to the $H \to ZZ$ decay mode.

On the contrary, the isotriplet model is severely constrained by its inherent
custodial symmetry breaking. With only 20 extra~fb of the cross section for the
most favourable value of $m_H = 300$~GeV, we have little hope to test this model
through double higgs production before we will reach the level of accuracy
that would allow us to test the SM directly.

However, further extension of the isotriplet model to the Georgi-Machacek model
changes the picture entirely. In this case the cross section can reach 2~pb, and
with the decays of the second Higgs boson to vector bosons possibly suppressed,
double higgs production might be the best mode to test this model at the LHC.

I would like to thank organizers for a warm and friendly atmosphere at the
conference, filled with sparkling magic of a picturesque Renaissance castle. I
am also grateful to my co-authors, S.~Godunov, A.~Rozanov and M.~Vysotsky.  I am
partially supported by grants RFBR No.~14-02-00995, NSh-3830.2014.2 and
MK-4234.2015.2.

\end{document}